\documentclass[preprint,showpacs,preprintnumbers,amsmath,amssymb,endfloats*]{revtex4}
\usepackage{graphicx}

\begin{document}

\thispagestyle{empty}
\title{
Comment on ``Effects of spatial dispersion on electromagnetic surface 
modes and on modes associated with a gap between two half spaces''
}
\author{
  G.~L.~Klimchitskaya${}^{1}$ and
V.~M.~Mostepanenko${}^{2}$
}

\affiliation{
${}^{1}$North-West Technical University, Millionnaya St. 5, St.Petersburg,
191065, Russia\\
${}^{2}$Noncommercial Partnership ``Scientific Instruments'', 
Tverskaya St. 11, Moscow, 103905, Russia
}

\begin{abstract}
Recently Bo E.~Sernelius [Phys. Rev. B {\bf 71}, 235114 (2005)] investigated
the effects of spatial dispersion on the thermal Casimir force between two
metal half spaces. He claims that incorporating spatial dispersion results
in a negligible contribution from the transverse electric mode at zero
frequency as compared to the transverse magnetic mode. We demonstrate
that this conclusion is not reliable
because, when applied to the Casimir effect, the 
approximate description 
of spatial dispersion used is unjustified. 
\end{abstract}


\maketitle

Ref.~\cite{1} is devoted to the question of how spatial dispersion affects the
Casimir force between real metal plates at nonzero temperature. As is
mentioned in Ref.~\cite{1}, there are two quite different predictions in
the literature regarding the thermal Casimir force between real metals.
According to one prediction \cite{2,3,4}, at separations of a few hundred
nanometers the thermal correction to the Casimir force is negligible
and agrees in a qualitative sense with the ideal metal result. At large
separations, where the magnitude of the total force is equal to the thermal
correction ($z>6\,\mu$m at $T=300\,$K), the results for real and ideal
metals coincide. According to another prediction \cite{5,5a}, at
short separations below one micrometer the thermal correction for real
metals can be as large as 10--20\% of the total Casimir force. At large
separations the thermal force between real metals is one-half of the
ideal-metal result. Mathematically the distinction between the two proposed
formalisms lies in the fact that in Refs.~\cite{2,3,4} there is a nonzero
contribution from the transverse electric mode at zero frequency, whereas
in Refs.~\cite{5,5a} this contribution is equal to zero. A nonzero result for
the contribution from the transverse electric mode at zero frequency is
obtained by using the dielectric function of the plasma model \cite{2,3}
or the Leontovich surface impedance \cite{4}. A zero result follows by 
application of the Drude dielectric function with a relaxation
parameter \cite{5,5a}. As was demonstrated in Refs.~\cite{6,7}, the
prediction of Refs.~\cite{5,5a} is excluded by experiment at 99\% confidence
within a wide separation region from 300  to 500\,nm,
whereas the predictions of Refs.~\cite{2,3,4} are consistent with
experiment. There is also the qualitative disagreement between the
prediction of Refs.~\cite{5,5a} and the results of the first modern
measurement of the Casimir force \cite{7a} at $z\approx 1\,\mu$m.
At $z= 1\,\mu$m the large thermal correction, predicted in Refs.~\cite{5,5a},
achieves 19\% of the force in ideal metal case, whereas the experimental
uncertainty in \cite{7a} was 5--10\%. The predicted thermal effect was not,
however, observed in \cite{7a}. 
Ref.~\cite{1} claims that the inclusion of spatial dispersion
leads to a result which is almost identical to that obtained from the
Drude model in Refs.~\cite{5,5a}. Below we demonstrate that including
spatial dispersion into the theory of the Casimir force as
in Ref.~\cite{1} is unjustified and, thus, the
result obtained in this reference cannot be trusted.

To find the electromagnetic surface modes associated with an empty gap
between the two half spaces, Ref.~\cite{1} assumes in Secs.~III and
IV the validity of the standard continuity boundary conditions
\begin{equation}
E_{1t}=E_{2t}, \quad
B_{1n}=B_{2n}, \quad
D_{1n}=D_{2n},\quad
B_{1t}=B_{2t}, 
\label{eq1}
\end{equation}
\noindent
where {\boldmath$n$} is the normal to the boundary directed
inside medium, and the subscripts $n,t$ refer to the normal
and tangential components, respectively. Here 
{\boldmath$E$} is the electric field strength, {\boldmath$D$}
is the electric displacement, {\boldmath$B$} is the magnetic
induction, and only nonmagnetic materials 
are considered so that for the magnetic field {\boldmath$H=B$}
(the subscript 1 refers to a vacuum and subscript
2 to a semispace). To describe the spatial dispersion, 
Ref.~\cite{1} assumes that the longitudinal and transverse
dielectric functions depend on both
the wave vector and frequency 
$\varepsilon_{kl}=\varepsilon_{kl}(\mbox{\boldmath$q$},\omega)$.
However, in the theory of the thermal Casimir force 
taking into account
spatial dispersion, both assumptions used in Ref.~\cite{1},
i.e., the boundary conditions (\ref{eq1}) and the dielectric
permittivities depending on both {\boldmath$q$} and $\omega$, are
unjustified.

While on the subject of boundary conditions, we recall that 
in Casimir-type problems the Maxwell equations of an electromagnetic
field in a metal can be written as
\begin{eqnarray}
&&\mbox{rot{\boldmath$E$}}+\frac{1}{c}
\frac{\partial\mbox{\boldmath$B$}}{\partial t}=0,
\quad
\mbox{div{\boldmath$D$}}=0,
\nonumber \\
&&\label{eqMax} \\
&&\mbox{rot{\boldmath$B$}}-\frac{1}{c}
\frac{\partial\mbox{\boldmath$D$}}{\partial t}=0,
\quad
\mbox{div{\boldmath$B$}}=0.
\nonumber
\end{eqnarray}
\noindent
Here it is assumed, that there are no current and charge densities
of the external sources (i.e., given independently of
{\boldmath$E,\> D$} and {\boldmath$B$}). The contribution from the
conduction electrons is taken into account by the relation for the
determination of {\boldmath$D$}
\begin{equation}
\frac{\partial\mbox{\boldmath$D$}}{\partial t}=
\frac{\partial\mbox{\boldmath$E$}}{\partial t}+
4\pi \mbox{\boldmath$i$},
\label{displacment}
\end{equation}
\noindent
where {\boldmath$i$} is the volume current density induced by
the fields {\boldmath$E$} and {\boldmath$B$}.

In electrodynamics with spatial dispersion the physical fields
{\boldmath$E$} and {\boldmath$B$} at the interface are
usually finite, though the displacement {\boldmath$D$} can
tend to infinity \cite{8}. Taking this fact into account,  
and integrating the Maxwell equations (\ref{eqMax})
over the thickness of the boundary layer
(see, e.g., Ref.~\cite{9}), one reproduces the first two
conditions of Eq.~(\ref{eq1}), but arrives to modified
third and fourth conditions \cite{8,8a}
\begin{eqnarray}
&&
E_{1t}=E_{2t}, \quad
B_{1n}=B_{2n}, 
\label{eq2} \\
&&
D_{2n}-D_{1n}=4\pi\sigma,\quad
[\mbox{\boldmath{$n$}}\times({\mbox{\boldmath{$B$}}}_2-
{\mbox{\boldmath{$B$}}}_1)]=\frac{4\pi}{c}\mbox{\boldmath{$j$}}.
\nonumber
\end{eqnarray}
\noindent
Here the induced surface charge and current densities are
given by
\begin{equation}
\sigma=\frac{1}{4\pi}\int_{1}^{2}
\mbox{div}[\mbox{\boldmath{$n$}}\times
[\mbox{\boldmath{$D$}}\times\mbox{\boldmath{$n$}}]]dl,
\quad
\mbox{\boldmath{$j$}}=\frac{1}{4\pi}\int_{1}^{2}
\frac{\partial \mbox{\boldmath{$D$}}}{\partial t}dl.
\label{eq3}
\end{equation}

In linear electrodynamics without spatial dispersion the
material equation connecting {\boldmath{$D$}} and
{\boldmath{$E$}} takes the form
\begin{equation}
{\mbox{{$D$}}}_{k}(\mbox{\boldmath$r$},t)=\int_{-\infty}^{t}
\hat{\varepsilon}_{kl}(\mbox{{\boldmath{$r$}}},t-t^{\prime})
\mbox{{{$E$}}}_{l}(\mbox{{\boldmath{$r$}}},t^{\prime})
dt^{\prime}
\label{eq4}
\end{equation}
\noindent
(here we consider a medium with time-independent
properties). According to Eq.~(\ref{eq4}), the electric 
displacement {\boldmath$D$}({\boldmath$r$},$t$) at a position
{\boldmath$r$} and time $t$ is determined by the electric
field {\boldmath$E$} at the same point {\boldmath$r$}
(the spatial dispersion is absent) but at different times
$t^{\prime}\leq t$ (generally speaking, the temporal dispersion
is present). It can be easily shown that the substitution of
Eq.~(\ref{eq4}) in Eq.~(\ref{eq3}) leads to $\sigma=0$,
 {\boldmath$j$}$=0$. In other words when only a temporal
dispersion is present the boundary conditions (\ref{eq2})
coincide with the standard continuity conditions (\ref{eq1}).
However, it is unjustified to use the boundary conditions
(\ref{eq1}) in the presence of spatial dispersion as is
done in Ref.~\cite{1}. Specific examples illustrate that in
this case neither $\sigma$ nor {\boldmath$j$} is equal to
zero \cite{8,10}. Note that the boundary conditions (\ref{eq2})
are the exact consequence of the Maxwell equations with sharp
boundary surfaces and they are obtained for macroscopic 
physical fields. It is not correct to mix them up with the
boundary conditions arising in perturbative theories and for
fictitious fields (see below).

We turn now to the second assumption used in Ref.~\cite{1},
i.e., to the description of spatial dispersion in terms of 
dielectric functions depending on  {\boldmath$q$} and
$\omega$. If only a temporal dispersion is present, one can 
substitute the Fourier transformations   
\begin{equation}
\mbox{{\boldmath{$E$}}}({\mbox{\boldmath$r$}},t)
=\int_{-\infty}^{\infty}
\mbox{\boldmath{$E$}}(\mbox{{\boldmath{$r$}}},\omega)
e^{-i\omega t}d\omega,
\quad
\mbox{{\boldmath{$D$}}}({\mbox{\boldmath$r$}},t)
=\int_{-\infty}^{\infty}
\mbox{\boldmath{$D$}}(\mbox{{\boldmath{$r$}}},\omega)
e^{-i\omega t}d\omega
\label{eq5}
\end{equation}
\noindent
in Eq.~(\ref{eq4}) and arrive at
\begin{equation}
\mbox{{{$D$}}}_{k}(\mbox{\boldmath$r$},\omega)=
\varepsilon_{kl}(\mbox{\boldmath$r$},\omega)
\mbox{{$E$}}_{l}(\mbox{{\boldmath{$r$}}},\omega),
\label{eq6}
\end{equation}
\noindent
where
\begin{equation}
\varepsilon_{kl}(\mbox{\boldmath$r$},\omega)=
\int_{0}^{\infty}
\hat{\varepsilon}_{kl}(\mbox{{\boldmath{$r$}}},\tau)
e^{i\omega\tau}d\tau
\label{eq7}
\end{equation}
\noindent
is the frequency-dependent dielectric permittivity and
$\tau\equiv t-t^{\prime}$. Eq.~(\ref{eq6}) and boundary
conditions (\ref{eq1}) are used in all traditional
derivations of the Lifshitz formula \cite{11,12,13,13a,14,15,16}
describing the van der Waals and Casimir force between parallel 
plates (half spaces) possessing only a temporal (in other
words ``frequency") dispersion.

If the material of the plates possesses both temporal and
spatial dispersion, Eq.~({\ref{eq4}) should be generalized to
\begin{equation}
\mbox{{{$D$}}}_{k}(\mbox{\boldmath$r$},t)=\int_{-\infty}^{t}
dt^{\prime}
\int d{\mbox{\boldmath$r$}}^{\prime}
\hat{\varepsilon}_{kl}(\mbox{{\boldmath{$r$}}},
{\mbox{\boldmath$r$}}^{\prime},t-t^{\prime})
\mbox{{{$E$}}}_{l}({\mbox{\boldmath{$r$}}}^{\prime},t^{\prime}).
\label{eq8}
\end{equation}
\noindent
If the medium were uniform in space (i.e., all points were
equivalent), the kernel $\hat{\varepsilon}$ would not depend 
on {\boldmath$r$} and {\boldmath$r$}${}^{\prime}$ separately, but
only on the difference 
{\boldmath$R\equiv r-r^{\prime}$}.
In this case, by performing the Fourier transformation
\begin{equation}
\mbox{{\boldmath{$E$}}}({\mbox{\boldmath$r$}},t)
=\int_{-\infty}^{\infty}d\omega\int d{\mbox{\boldmath$q$}}
\mbox{\boldmath{$E$}}(\mbox{{\boldmath{$q$}}},\omega)
e^{i({\mbox{\scriptsize\boldmath$qr$}}-\omega t)}
\label{eq9}
\end{equation}
\noindent
(and the same for {\boldmath$D$}), and substituting it in
Eq.~(\ref{eq8}), one could introduce the dielectric permittivities
\begin{equation}
\varepsilon_{kl}(\mbox{\boldmath$q$},\omega)=
\int_{0}^{\infty}d\tau\int d\mbox{\boldmath$R$}\,
\hat{\varepsilon}_{kl}(\mbox{{\boldmath{$R$}}},\tau)
e^{-i(\mbox{\scriptsize\boldmath$qR$}-\omega\tau)}.
\label{eq10}
\end{equation}
\noindent
These permittivities depend on both the wave vector and frequency 
and bring Eq.~(\ref{eq8}) into the form
\begin{equation}
\mbox{{{$D$}}}_{k}(\mbox{\boldmath$q$},\omega)=
\varepsilon_{kl}(\mbox{\boldmath$q$},\omega)
\mbox{{$E$}}_{l}(\mbox{{\boldmath{$q$}}},\omega)
\label{eq11}
\end{equation}
\noindent
in analogy to Eq.~(\ref{eq6}).

In Ref.~\cite{1}, however, the system under consideration is
not uniform due to the presence of a macroscopic gap between the
two half spaces. Because of this, it is impossible to assume
that the kernel $\hat{\varepsilon}$ depends only on
{\boldmath$R$} and $\tau$, and hence it is not possible to introduce
$\varepsilon_{kl}(\mbox{\boldmath$q$},\omega)$. In fact in the presence
of boundaries the kernel $\hat{\varepsilon}$ for systems with
spatial dispersion depends not only on  {\boldmath$R$} and $\tau$ 
but also on the distance from the boundary \cite{8}.
An approximate phenomenological
approach to dealing with this situation is described
in Ref.~\cite{8}. For electromagnetic waves with
a wavelength $\lambda$ the kernel
$\hat{\varepsilon}(\mbox{\boldmath$r$},
\mbox{\boldmath$r$}^{\prime},\tau)$ is significantly large only
in a certain vicinity of the point {\boldmath$r$} with
characteristic dimensions $a\ll\lambda$ (in fact 
for nonmetallic condensed media
$a$ is of the order of the lattice constant). 
One can then assume that $\hat{\varepsilon}$ is a function of 
{\boldmath$R$}$=${\boldmath$r$}--{\boldmath$r$}${}^{\prime}$,
except for a layer of thickness $a$ adjacent to the
boundary surface. If one is not interested in this
subsurface layer, the quantity 
$\varepsilon_{kl}(\mbox{\boldmath$q$},\omega)$ may be used
to describe the remainder of the medium. 

This approximate phenomenological approach is widely applied 
in the theory of the anomalous skin effect (see, e.g., 
Ref.~\cite{Kli}) for the investigation of bulk physical phenomena
involving electromagnetic fields. In Ref.~\cite{Kli} some
kind of fictitious infinite system is introduced and the electromagnetic
fields in this system are discontinuous on the surface. This
discontinuity should not be confused with the discontinuity of
physical fields of a real system in the presence of spatial
dispersion given by Eqs.~(\ref{eq2}) and  (\ref{eq3}).
(There is, however,
another approach which describes the anomalous skin effect
in polycrystals in terms of the local Leontovich impedance
taking into account the shape of Fermi surface \cite{Kag}).
The frequency and wave vector dependent dielectric permittivity
in the presence of boundaries is also successfully applied in
some other problems, e.g., in the theory of radiative heat
transfer \cite{21a} or in the study of electromagnetic interactions
of molecules with metal surfaces \cite{21b}. All these applications
are in more or less good agreement with experiment.
The boundary effects can be taken
into account by the boundary conditions (\ref{eq2})
supplemented in some way by so called ``additional
boundary conditions".

It is unlikely, however, that the phenomenological
approach of Ref.~\cite{Kli} using the frequency and wave vector 
dependent dielectric permittivity
would be applicable for the calculation of Casimir
force between metal surfaces, where the boundary effects
for vacuum electromagnetic oscillations
are of prime importance and contribute significantly to
the total result. Ref.~\cite{1} uses the
quantities $\varepsilon_{kl}(\mbox{\boldmath$q$},\omega)$ in a
spatially nonuniform system of two half spaces
separated by a gap without justification, and determines the
surface modes from the continuity conditions (\ref{eq1})
instead of the boundary conditions (\ref{eq2}) valid in
the presence of spatial dispersion. 
As was emphasized in Ref.~\cite{8}, in the case of a bounded
medium it is inadmissible to use the same kernel in Eq.~(\ref{eq8})
which is appropriate for an unbounded medium. 
According to Ref.~\cite{17}, this leads to a
violation of the law of conservation of energy. 
It is then not surprising that in doing so Ref.~\cite{1} arrives
to an incorrect contribution from the transverse electric mode
to the Casimir free energy.
However, in addition to
Ref.~\cite{1} some other recent publications (see, e.g.,
Refs.~\cite{18,19}) apply the dielectric permittivities
$\varepsilon_{kl}(\mbox{\boldmath$q$},\omega)$ obtained for an unbounded 
medium to the theory of the Casimir effect which essentially depends 
on the presence of boundaries.

One more shortcoming of Ref.~\cite{1} is that it uses the
usual Lifshitz formula for the free energy derived in the
presence of only temporal dispersion \cite{11,12,13,13a,14,15,16},
to compute the effects of spatial dispersion. It has been 
known, however, that with the inclusion of spatial dispersion 
the free energy of a fluctuating field takes the form
${\cal F}={\cal F}_L+\Delta{\cal F}$, where
${\cal F}_L$ is given by the Lifshitz formula derived in a spatially
local case and written in terms of the Fresnel reflection
coefficients, and
$\Delta{\cal F}$ is an additional term which can be expressed
in terms of the thermal Green's function of electromagnetic
field and polarization operator \cite{20}. The review paper
\cite{20} contains a few references to incorrect results by
different authors obtained by the substitution of
$\varepsilon_{kl}(\mbox{\boldmath$q$},\omega)$, 
taking account of spatial dispersion,
into the usual Lifshitz formula. During the last years
several more similar papers have been published
(see, e.g., Refs.~\cite{18,19}). It can be true that the Lifshitz
formula written in terms of general reflection coefficients is
applicable in both spatially local and nonlocal situations.
However, as far as the exact reflection coefficients in a
spatially nonlocal case are unknown, the use of some approximate
models, elaborated in the literature for 
applications different than the Casimir effect,
may lead to incorrect results for $\Delta{\cal F}$ and create
serious inconsistencies with experiment.

To conclude, the main purpose of Ref.~\cite{1}
``to find out how spatial dispersion affects the Casimir
force between real metal plates" is in our opinion not
achieved. Instead of using the boundary conditions (\ref{eq2})
valid in the presence of spatial dispersion, and the
appropriate generalization of the Lifshitz formula,
Ref.~\cite{1} uses the continuity conditions (\ref{eq1})
and usual Lifshitz formula derived in the presence
of only temporal dispersion. Ref.~\cite{1} also applies
the phenomenological dielectric permittivity which depends 
on frequency and wave vector to the calculation of the
Casimir force, a phenomenon primarily determined by virtual 
photons. These unjustified assumptions are responsible for
the main conclusion of Ref.~\cite{1} that spatial
dispersion results in practically the same ``dramatic
effect" that was previously obtained in Ref.~\cite{5}
from the Drude dielectric function: the contribution
from the transverse electric mode at zero frequency
``completely vanishes if dissipation is included",
and is negligibly small compared to the contribution
from the transverse magnetic mode if dissipation is
absent. Ref.~\cite{1} does not inform the reader that
theoretical approaches with zero or negligibly small
contribution from the transverse electric mode at zero 
frequency have already been excluded experimentally at 
99\% confidence in Refs.~\cite{6,7}.
By now such approaches are rejected by 4 different
experiments \cite{25a} including the experiment of
Ref.~\cite{7a}.
At the same time, the Leontovich impedance approach, which
leads to a nonzero contribution from the transverse electric 
mode at zero frequency, is both consistent with experiment
and takes dissipation into account \cite{4,6,7,20a}.
The physical reasons why the Leontovich impedance approach
is to be preferred are articulated in Ref.~\cite{20a}.

As to the role of spatial dispersion in the Casimir
effect, it was first investigated using the Leontovich
surface impedance in Ref.~\cite{21}. 
The Leontovich approach can be considered as an approximate
phenomenological approach. As was shown above, the approach
of Ref~\cite{1}, in spite of the author's claims to the
contrary, is also phenomenological and approximative.
At the moment
there is no fundamental theory which describes the thermal
Casimir force between real metals incorporating
spatial nonlocality.
Because of this, one should choose between different
approximate aproaches, and in so doing the role of experiment 
must not be underestimated. It is well understood
that at laboratory temperatures of about 300\,K and at 
Casimir plate separations of a
few hundred nanometers (the frequency region of infrared
optics) spatial dispersion does not play any role. 
It can be confidently neglected also in the region of the 
normal skin effect, i.e., at separations between the plates
greater than 4--5\,$\mu$m. If some approximate phenomenological
approach leads to the opposite conclusion, there are strong
reasons to believe that it is not reliable.

\section*{Acknowledgments}
This work was supported by the Russian Foundation for
Basic Research (Grant No. 05--08--18119a).

\end{document}